\documentclass{PoS}
\usepackage{slashed}
\usepackage{graphicx}
\usepackage{color}

\title{$D_s$ to $\phi$ and other transitions from lattice QCD}

\ShortTitle{$D_s$ to $\phi$ and other transitions from lattice QCD}

\author{\speaker{Gordon Donald}\\
        SUPA, School of Physics and Astronomy, University of Glasgow, Glasgow, G12 8QQ, UK\\
	School of Mathematics, Trinity College, Dublin 2, Ireland\\
        E-mail: \email{donaldg@tcd.ie}} %

\author{Christine Davies\\
        SUPA, School of Physics and Astronomy, University of Glasgow, Glasgow, G12 8QQ, UK}
\author{Jonna Koponen\\
        SUPA, School of Physics and Astronomy, University of Glasgow, Glasgow, G12 8QQ, UK \\}
\author{HPQCD Collaboration\\}

\abstract{We have studied transitions between vector and pseudoscalar mesons using the HISQ action for the valence quarks.
We have calculated all of the axial and vector form factors that appear in the decay rate for $D_s \to \phi \ell \nu$ over the full $q^2$ range and compared them to the shape of the experimental decay distributions. 
We use nonperturbatively normalised currents for the vector and axial vector currents. 
The same set up for the three point correlations functions also allow us to study radiative decays and we have calculated the decay rate for $J/\psi \to \eta_c \gamma$.
}

\FullConference{31st International Symposium on Lattice Field Theory LATTICE 2013\\
		 July 29 -- August 3, 2013\\
		 Mainz, Germany}

\begin{document}

\section{The semileptonic decay $D_s \to \phi \ell \nu$}
\subsection{Lattice calculation}

We have calculated the vector and axial vector form factors which appear in the decay $D_s \to \phi \ell \nu$.
This is a transition between pseudoscalar and vector mesons and the same methods can be used for other transitions, such as the charmonium radiative decay $J/\psi \to \eta_c \gamma$ \cite{charmpaper}.

We use the improved staggered HISQ action \cite{hisq} for $s$ and $c$ valence quarks and work on gauge configurations generated by the MILC Collaboration that include $ 2+1$ flavours of asqtad sea quarks \cite{milcasqtad}.
The details of the configurations we have used are listed in Table \ref{confs}.

\begin{table}
\begin{center}
\begin{tabular}{llllll}
\hline
\hline
Set &  $~a/fm$ & $au_0m_{l}^{asq}$ / $au_0m_{s}^{asq}$ & $L_s/a \times L_t/a$ & $n_{cfg}$ & $T$ \\
\hline
1 &  0.12 & 0.005/0.05 & 24 $\times$ 64 & 2088 & 12, 15, 18 \\
2 &  0.12 & 0.01/0.05 & 20 $\times$ 64 & 2259 & 12, 15, 18 \\
\hline
3 &  0.09 & 0.0062/0.031 & 28 $\times$ 96 & 1911 & 16, 19, 20, 23\\
\hline 
\hline
\end{tabular}
\caption{Details of the gauge configurations used in this calculation. The configurations were generated by the MILC collaboration and contain $N_f = 2+1$ flavours of asqtad sea quarks.}
\label{confs}
\end{center}
\end{table}

\begin{figure}
\begin{center}
\includegraphics[angle=0,width=6cm]{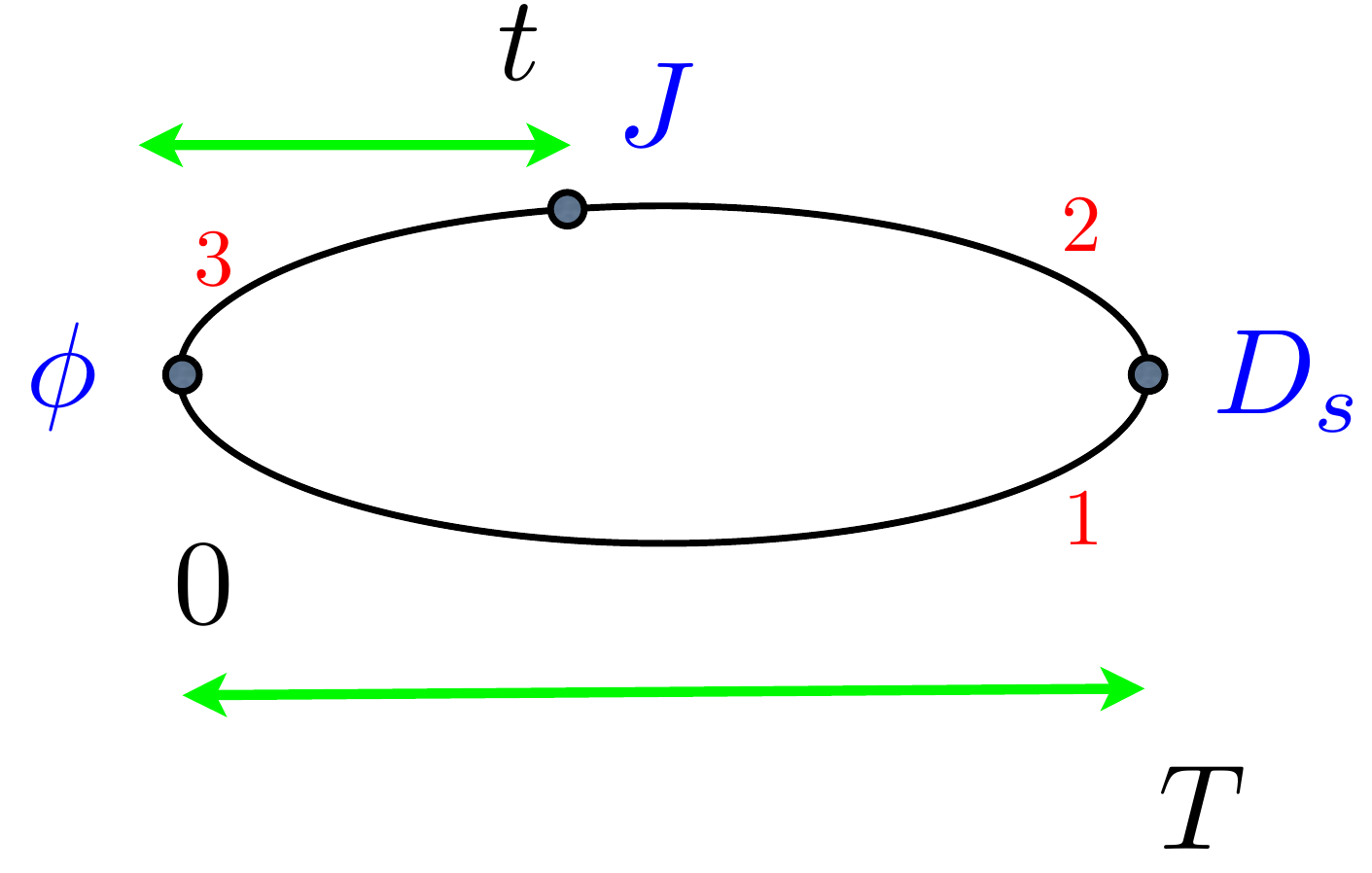}
\caption{A diagram of the quark propagators in the 3-point function. Propagators 1 and 3 are strange quark propagators and propagator 2 is a charm propagator.}
\label{3pt}
\end{center}
\end{figure}

To extract the transition matrix element, we calculate 2-point and 3-point correlation functions.
We simulate the $\phi$ meson as a $\bar{s}s$ vector meson and a diagram of the 3-point correlator for the $D_s \to \phi$ transition is shown in Figure \ref{3pt}.
We create a $\phi$ meson at time $0$ and destroy the $D_s$ at $T$. At an intermediate time $t$, we insert a vector or axial vector current.
We perform a simultaneous fit to the $D_s$ and $\phi$ 2-point correlators and 3-point correlators with multiple values of $T$.

We use Bayesian methods \cite{gplbayes} to fit the correlators to the following fit forms where we include oscillating contributions from opposite parity states because we are using staggered quarks:
\begin{equation}C^{(P)}_{2pt}(t) = \sum_{i} \{a^{(P)}_{i}\}^2 [e^{-Et} + e^{-E(L_t-t)}] +\mbox{oscillations} \end{equation}
\begin{equation}
C^{P\rightarrow Q}_{3pt}(t,T) =  \sum_{i,j} a^{(P)}_{i} [e^{-Et} + e^{-E(L_t-t)}] J_{i,j} a^{(Q)}_{j} [e^{-E(T-t)} + e^{-E(L_t-T+t)}] +\mbox{oscillations}. 
\end{equation}

The fit parameters $a_i$ appear in both the 2-point and 3-point fit forms and this allows us to extract the $J_{i,j}$ which are simply related to the current matrix elements we need, $\langle D_s| \Gamma|\phi\rangle$.

The general form for the $D_s \to \phi$ transition matrix element is given by \cite{richmanburchat}
\begin{eqnarray} \langle \phi(p', \varepsilon) | V^\mu - A^\mu | D_s (p) \rangle & = & \frac{2i\epsilon^{\mu\nu\alpha\beta}}{m_{D_s}+m_{\phi}} \varepsilon_\nu p_\alpha p'_\beta V(q^2) - (m_{D_s}+m_\phi)\varepsilon^\mu A_1(q^2) \label{me}
\\ & + & \frac{\varepsilon \cdot q}{m_{D_s}+m_\phi}(p+p')^\mu A_2(q^2) + 2m_\phi \frac{\varepsilon \cdot q}{q^2} q^\mu (A_3(q^2) - A_0(q^2)) \nonumber
\end{eqnarray}
where the $A_3(q^2)$ form factor is defined as
\begin{equation} A_3(q^2) = \frac{m_{D_s}+m_\phi}{2m_\phi} A_1(q^2) - \frac{m_{D_s}-m_\phi}{2m_\phi} A_2(q^2). \end{equation}
At $q^2 = 0$, the form factors obey the relation \begin{equation} A_3(0) = A_0(0). \end{equation}


We simulate with the $D_s$ meson at rest and access the full physical range of $q^2$ (from $q^2 = 0$ to $q^2_{max} = (m_{D_s}-m_\phi)^2$) by giving momentum to the $\phi$ meson.
We use twisted boundary conditions \cite{twistedbcs} to tune $p_\phi$ and extract each form factor from Equation \ref{me} by arranging the kinematics such that the expression for the matrix element simplifies.

We can get $A_1(q^2)$ by choosing the $\phi$ polarisation and momentum so that $\varepsilon \cdot q =0$. 
In this case, Equation \ref{me} reduces to 
\begin{equation}\langle \phi(p', \varepsilon) | A^\mu | D_s (p) \rangle = (m_{D_s}+m_\phi)\varepsilon^\mu A_1(q^2). \end{equation}

We have some choice of the staggered $\phi$ and $A_\mu$ operators we use in the correlators as we can use local or point-split operators for $A_\mu$ and $\phi$.
These operators differ only in taste so the differences between them are a discretisation effect.
We find for HISQ that these are very small.
The vector and axial vector operators we use for the weak currents are nonperturbatively normalised.
Details of the methods used for current normalisation are in \cite{gdlatt11}.
The $A_0(q^2)$ form factor is not experimentally accessible as it is suppressed by the lepton mass, but is straightforward to calculate.
Using the PCAC relation, we have
\begin{equation} q_\mu \langle \phi(p', \varepsilon) | A^\mu | D_s (p) \rangle = 2m_\phi \varepsilon \cdot q A_0(q^2) = (m_c + m_s) \langle \phi(p', \varepsilon) | \gamma_5 | D_s (p) \rangle. 
\end{equation}
The local pseudoscalar current needed here is absolutely normalised.

The $A_2(q^2)$ form factor is needed to reconstruct the decay rate, but only enters the matrix element when $\varepsilon \cdot q \neq 0$.
The easiest way to extract it from the matrix element is to use our knowledge of $A_1(q^2)$ and $A_0(q^2)$.
This is easiest at $q^2=0$ because we have relation $A_0(0) = A_3(0)$.

We get $V(q^2)$ using a vector current insertion in our 3-point correlator.
To make a taste-singlet correlator, we use the spin-taste $\gamma_\mu \otimes \gamma_\mu\gamma_\nu$ vector operator for our $\phi$ and use $\gamma_t\gamma_5 \otimes \gamma_t\gamma_5$ for the $D_s$.
For a vector current, Equation \ref{me} becomes
\begin{equation} \langle \phi(p', \varepsilon) | V_\alpha | D_s (p) \rangle = \frac{2i\epsilon_{\mu\nu\alpha t}}{m_{D_s}+m_{\phi}} \varepsilon^\mu E_{D_s} p_\phi^\nu V(q^2). \label{vff}\end{equation}

\begin{figure}
\begin{center}
\includegraphics[angle=-90,width=0.8\textwidth]{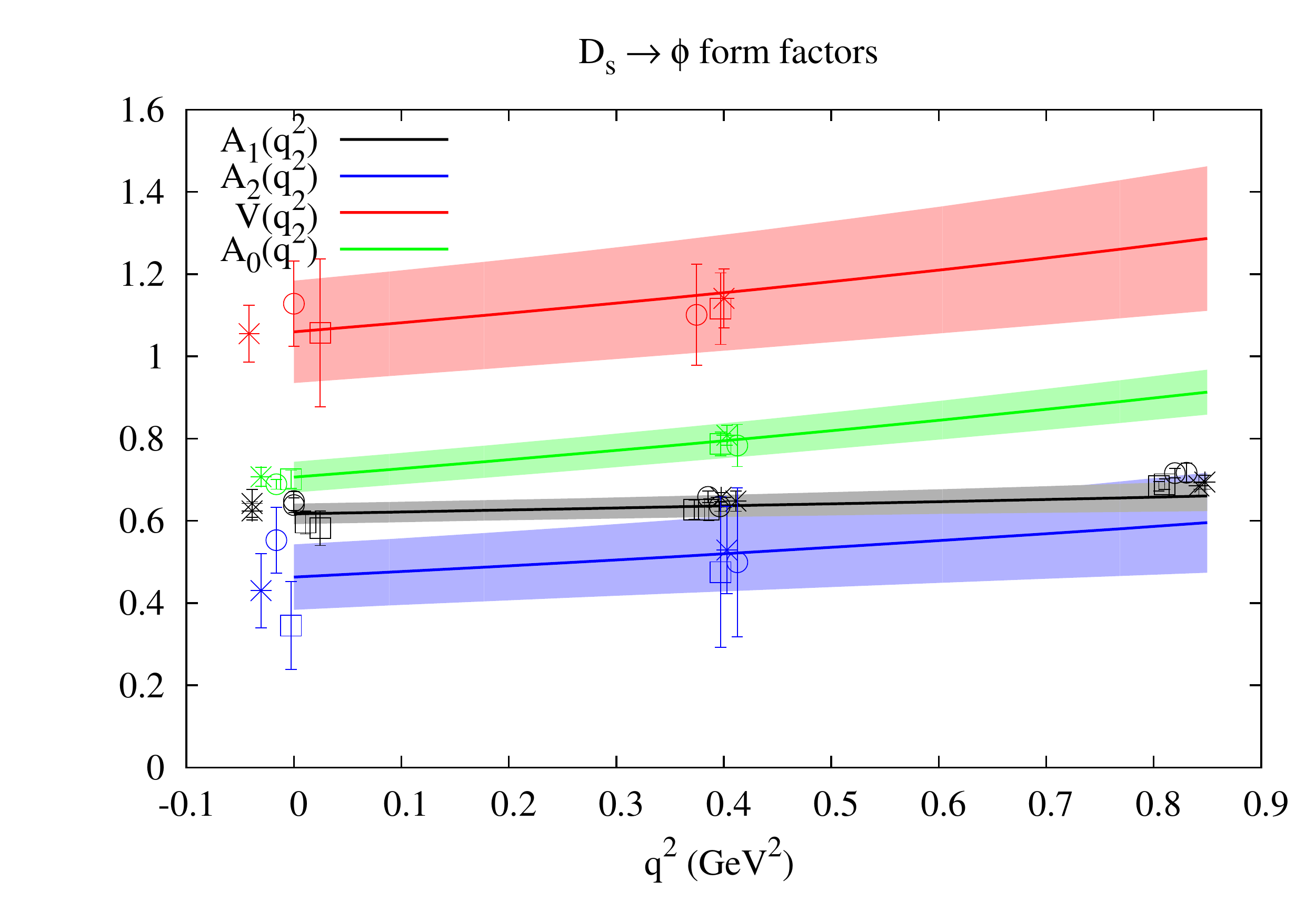}
\caption{The $D_s \to \phi \ell \nu$ form factors as a function of $q^2$. The points are the lattice data on each configuration and the shaded bands are the extrapolated form factors between $q^2=0$ and $q^2_{max}$.}
\label{ff}
\end{center}
\end{figure}

We plot the form factors we obtain in Fig \ref{ff}. The points are the lattice data on each of the three ensembles we use and the shaded bands show the form factors as a function of $q^2$ for the physical range after extrapolating to the continuum and to physical sea quark masses.

The physical limit is reached using an extrapolation in $z$ space, defined as

\begin{equation} z(q^2) = \frac{\sqrt{t_+ - q^2} - \sqrt{t_+} }{\sqrt{t_+ - q^2} + \sqrt{t_+} }, \end{equation}
where $t_+ = (M_{D_s} + M_\phi)^2$ and $t_0 = 0$.

By converting $q^2$ to $z$, the semileptonic region is mapped to $|z|<1$ and the form factor can be fitted as a power series in $z$.
We do the chiral and continuum extrapolation by allowing the co-efficients of each term in the z-expansion to depend on lattice spacing and the difference between the sea quark masses and their physical values, $\delta_l$ \cite{hna}.
For each form factor ($A_1$, $A_2$, $V$, $A_0$), we fit to
\begin{equation} \tilde{F}(z) = \sum^3_{n=0} A_n\left\{ 1 + B_na^2 + C_na^4 + D_n\delta_l \right\} z^n. \end{equation}

\subsection{Comparison with experiment}

The differential decay rate for $D_s \to \phi \ell \nu$ is given by \cite{richmanburchat}
\begin{eqnarray}
&\frac{d\Gamma(D_s \rightarrow \phi \ell \nu, \phi \rightarrow K^+K^-)}{dq^2d\cos \theta_K d\cos \theta_{\ell} d\chi} =&\frac{3}{8(4\pi)^4}G_F^2|V_{cs}|^2\frac{p_{\phi}q^2}{M^2_{D_s}}{\mathcal{B}}(\phi \rightarrow K^+K^-) \times \\
&&\left\{ (1+ \cos \theta_{\ell})^2\sin^2 \theta_K |H_+(q^2)|^2 \right.\nonumber \\ 
&& + (1- \cos \theta_{\ell})^2\sin^2 \theta_K |H_-(q^2)|^2 \nonumber \\ 
&& + 4 \sin^2\theta_{\ell} \cos^2 \theta_K |H_0(q^2)|^2 \nonumber \\
&& + 4 \sin \theta_{\ell} (1+\cos\theta_{\ell})\sin \theta_K \cos \theta_K\cos \chi H_+(q^2)H_0(q^2) \nonumber \\
&& - 4 \sin \theta_{\ell} (1-\cos\theta_{\ell})\sin \theta_K \cos \theta_K\cos \chi H_-(q^2)H_0(q^2) \nonumber \\
&& \left. - 2 \sin^2 \theta_{\ell} \sin^2 \theta_K \cos 2 \chi H_+(q^2)H_-(q^2) \right\} . \nonumber 
\end{eqnarray}

\begin{figure}[h]
\begin{center}
\includegraphics[angle=0,width=.5\textwidth]{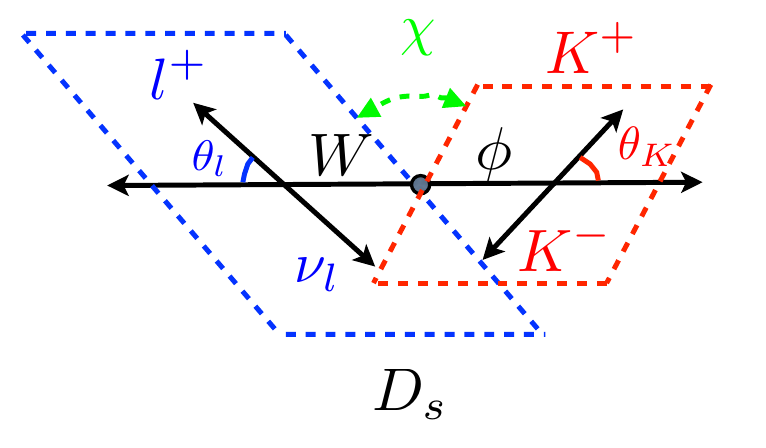}
\caption{A diagram of the angles which appear in the decay rate. }
\label{angles}
\end{center}
\end{figure}

The decay angles are defined for a $D_s^+$ decay as: $\theta_{\ell(K)}$ the angle between the momentum of the $\ell(K^+)$ and the centre of mass momentum of the $\ell \nu (K^+K^-)$ pair, and $\chi$ is the angle between the two planes defined by the $\ell\nu$ and $K^+K^-$ pairs. For a $D_s^-$, the $K^-$ is used in place of the $K^+$ in the definition of $\theta_K$ and $\chi \to -\chi$.
The decay angles are shown in Figure \ref{angles}.

The decay distribution is written in terms of helicity amplitudes (as the $D_s$ is a pseudoscalar, the $W$ and $\phi$ helicities are constrained to be the same), which are related to the form factors as
\begin{equation}H_{\pm}(q^2) = (M_{D_s}+M_{\phi})A_1(q^2) \mp \frac{2M_{D_s}p_{\phi}}{M_{D_s}+M_{\phi}}V(q^2).\end{equation}
and
\begin{equation}
H_0(q^2) = \frac{1}{2M_{\phi}\sqrt{q^2}} [ (M_{D_s}^2-M_{\phi}^2-q^2)(M_{D_s}+M_{\phi})A_1(q^2) 
- 4 \frac{M_{D_s}^2p_{\phi}^2}{M_{D_s}+M_{\phi}}A_2(q^2) ] .
\end{equation}

\begin{figure}
\begin{center}
\includegraphics[angle=-90,width=.87\textwidth]{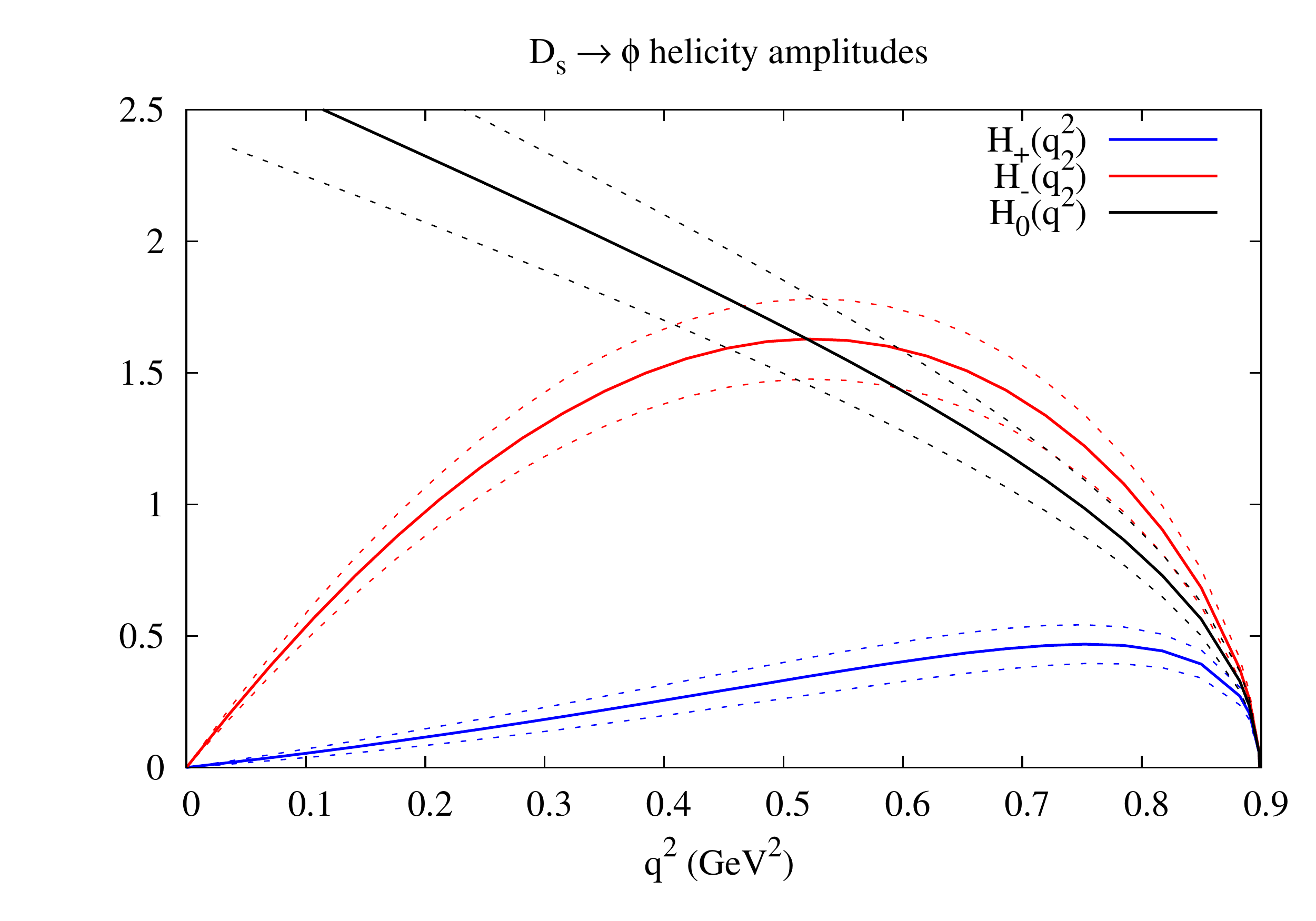}
\caption{Each of the $D_s \to \phi \ell \nu$ helicity amplitudes as a function of $q^2$. The helicity amplitudes $H_i(q^2)$ are plotted as $p_\phi q^2 |H_i(q^2)|^2$. At $q^2_{max}$, $p_\phi = 0$ so all the helicity amplitudes vanish.}
\label{ha}
\end{center}
\end{figure}

We construct these helicity amplitudes from the continuum extrapolated form factors.
In Figure \ref{ha}, we plot the the combination $p_\phi q^2 |H_i(q^2)|^2$, which is how they appear in decay rate.
After integrating over the angular distributions, each helicity amplitude has the same overall factor in the decay rate, so the relative contribution of each one to the distribution in $q^2$ is as shown in Figure \ref{ha}.

\begin{figure}
\begin{center}
\includegraphics[angle=-90,width=.87\textwidth]{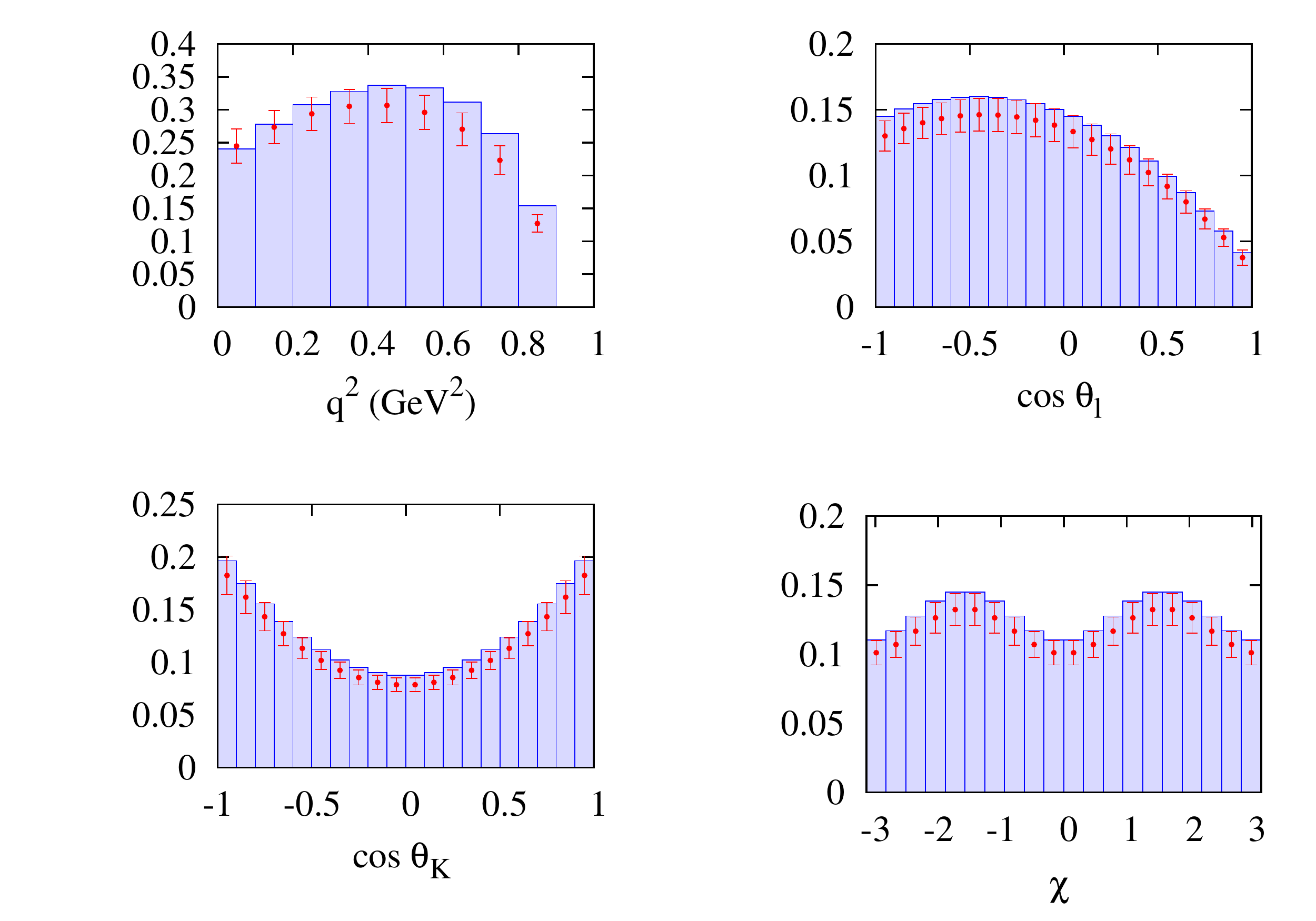}
\caption{The decay distributions as a function of each of the kinematic variables. The red points are obtained by integrating the lattice form factors and including the value of $V_{cs}$ from unitarity and the blue bins are from BaBar.} 
\label{dists}
\end{center}
\end{figure}

We can reconstruct the decay distributions as functions of all four of the kinematic variables by integrating over the other three. 
The angular integrals are straightforward and we do the $q^2$ integration numerically.
These are plotted in Figure \ref{dists}, where the red points (with error bars) are the decay rate for each bin calculated from our lattice form factors.
The blue histograms are the decay distributions from the form factors obtained by the BaBar experiment \cite{babar}. 

We can extract a value for $V_{cs}$, the CKM matrix element which appears in the charm to strange decay.
By integrating over all the kinematic variables, we can obtain a lattice determination of the total decay rate and compare to BaBar's measurement of the $D_s \to \phi \ell \nu$ branching fraction.
The lattice and experimental decay rates differ by a factor of $V_{cs}^2$.
This comparison gives us $V_{cs} = 1.017(60)$, which is in agreement with values of $V_{cs}$ extracted from lattice calculations of pseudoscalar to pseudoscalar semileptonic decays \cite{dtok} and the $D_s$ leptonic decay rate \cite{fds}.
It also agrees with $V_{cs}$ from unitarity.

Our final value for $V_{cs}$ includes a systematic error to account for the coupling of the $\phi$ meson to $K\bar{K}$ states \cite{dsphipaper}.

\section{Other transitions}
We can use similar staggered 3-point correlation functions to study other transitions. 
For example, the same methods used to find the vector form factor can be used for the  $J/\psi \to \gamma \eta_c$ radiative decay, as studied in \cite{charmpaper}.
Similar correlation functions are required for the decay of a pseudoscalar meson $\pi^0 / \eta_c \to \gamma\gamma$.

Further semileptonic transitions between pseudoscalar and vector mesons can be studied using heavier quark masses than charm.
For example, these methods could be used to extract form factors for $B_{(s)} \to D^*_{(s)} \ell \nu$. 



\section*{Acknowledgements}
We are grateful to MILC for the use of their gauge configurations. We used the Darwin Supercomputer as part of the DiRAC facility jointly funded by STFC, BIS and the Universities of Cambridge and Glasgow. This work was funded by STFC.

\end{document}